# Phase Drift Compensating RF Link for Femtosecond Synchronization of E-XFEL.

Dominik Sikora, K. Czuba, *Member, IEEE*, P. Jatczak, M. Urbanski, H. Schlarb, F. Ludwig and H. Pryschelski

*Abstract*—Modern high-energy particle accelerators and Free-Electron Lasers incorporate large quantities of sensitive RF and microwave frequency devices distributed over kilometer distances. Such devices require extreme stable phase and time synchronization by means of high frequency signal distributed along the accelerator facility.

Coaxial cables are commonly used to distribute the reference signal over the large machine to synchronize electronic systems and they are the main source of undesirable phase drifts in the synchronization system. Signal phase drifts in cables are mainly caused by temperature and humidity variations and their values usually exceed required phase synchronization accuracy by more than order of magnitude.

There are several approaches to reduce signal phase drifts in coaxial cables. This paper describes the realization of active phase stabilization system based on interference phenomenon. A phase-locked signal from the transmitter is reflected at the end of a coaxial cable link. Directional couplers placed along the cable pick up the forward and reflected signals and interfere them to cancel out the cable phase drifts. Distributed hardware including interferometer controller/transmitter and receiver modules were built to demonstrate system concept and performance. Link input and output devices used FPGA I/O boards with Ethernet interface to control system operation. Specialized firmware and software was developed to calibrate and control the system.

This paper describes the concept of interferometer link, designed hardware, basic control algorithms and performance evaluation results. The link prototype was built to distribute 1.3 GHz signal through a coaxial cable. Measured phase drift suppression factor value exceeded level of 100.

## I. Introduction

EROPEAN XFEL which is a modern Free Electron Laser facility [1] requires extreme stable phase and time synchronization [2]. A long term phase stability of the RF synchronization system has to be in a femtosecond scale. Coaxial cables, which are commonly used as a distribution media for the phase reference signal are the main source of phase drifts [3] and their values usually exceed requirements by more than two orders of magnitude. This leads to necessity of using a reference signal distribution system with an active phase drift compensation.

The phase averaging reference line concept [4-7] shows that phase drifts from coaxial lines in synchronization systems can be significantly suppressed. The main disadvantage of the concept is manual and very precise work that has to be done to set the link up. Another drawback is that there is no control on the link settings after starting of its operation. There are factors (e.g. aging, temperature and humidity variations) that could influence on the link parameters and thus, on the operating point, which should be set very precisely and maintained unchanged during entire operation of the link. Component parameter changes can simply decrease the link performance which can be represented by a phase drift Suppression Factor (SF), which is defined as a ratio of unsuppressed to suppressed phase drift values.

The interferometer link concept was developed to provide fully automatic adjustments of parameters and more reliable operation. It was very desirable for the E-XFEL because all synchronization system components were supposed to be placed in the tunnel with very limited access. The new concept allows to set the interferometer link to an operating point with a maximum suppression factor and to adjust it in any time.

## II. Interferometer Link Concept

A simplified schematic of the new link concept is shown in Fig. 1 a). The interferometer link consist of a transmitter called an InCon (Interferometer Controller) and a receiver called a TapPoint. A phase reference signal from the source is distributed forth and back through the InCon (as a forward signal), the Cable 1, the TapPoint and the Cable 2 (main coaxial cable) to the GND short at InCon. Signal reflected from the GND short goes back to the TapPoint, where fractions of the forward and reflected signals are coupled out from the main line. Both signals interfere there in an RF power combiner to cancel out the phase drifts appearing in the main coaxial cable (between InCon and TapPoint).

This concept requires fulfilling three basic conditions to achieve large (few hundred) SF values:
1.  Phase lock of the signal at the GND short to the reference signal source – performed by a Phase-Lock

This paragraph of the first footnote will contain the date on which you submitted your paper for review. This work was supported by Polish Ministry of Science and Higher Education, founds for international co-finance projects for year 2017.

D. Sikora, K. Czuba, P. Jatczak and M. Urbanski are with the Warsaw University of Technology, Institute of Electronic Systems, 00-665 Warsaw, Poland (e-mail: D.Sikora@elka.pw.edu.pl).

H. Schlarb, F. Ludwig and H. Pryschelski are with the Deutsches Elektronen-Synchrotron DESY, 22607 Hamburg, Germany.

2. Equal amplitudes and phase shifts against the signal source of forward and reflected signals at the TapPoint's RF combiner. They are adjusted by a variable attenuator Att2 and a variable phase shifter PhS4.

Loop in the InCon,

3. Appropriate electrical length between InCon and TapPoint's directional couplers (C4 and C5) to minimize influence of link error signals.– It can be adjusted by the main line phase shifter PhS2 in the TapPoint.

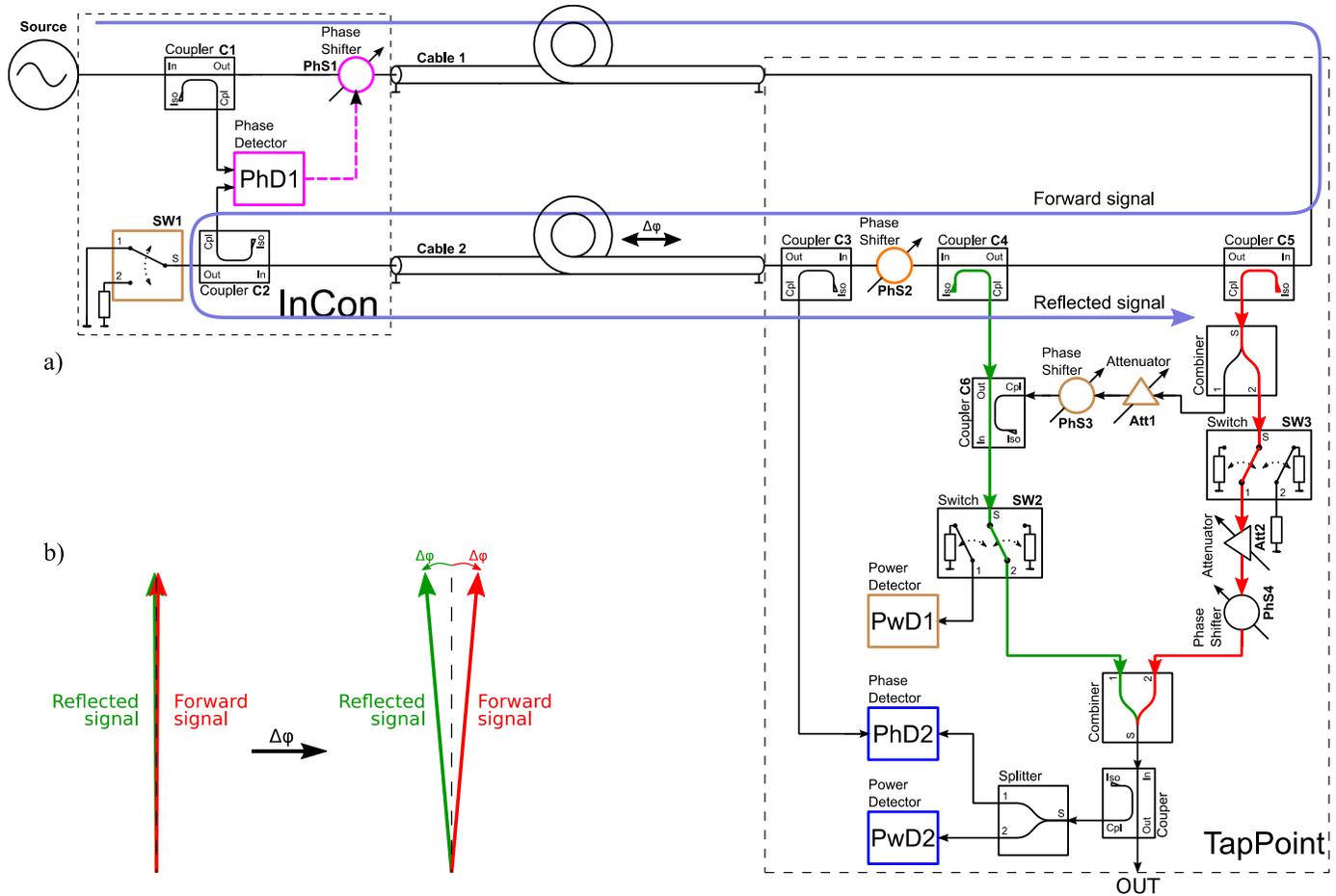

Fig. 1. a) Simplified schematic of the interferometer RF link,  b) Basic behavior of forward and reflected signals at the link output.

A basic signal flow in the TapPoint and behavior of the signals at the TapPoint's RF combiner is shown in Fig. 1. Directional couplers C4 and C5 pick up parts of the forward and reflected signals into the TapPoint circuits. Variable attenuator Att2 and variable phase shifter PhS4 are used to adjust both signals phases and amplitudes to fulfil the 2$^{nd}$ basic condition (see Fig. 1 b) left). Adjusted signals are summed in the RF combiner. The output signal has two times higher amplitude and the same phase as the combined forward and reflected signals. The first basic condition ensures that any phase drift $\Delta\varphi$ appearing in the Cable 2 changes the phase of the forward signal by the same value but in opposite direction than the phase of the reflected signal (-$\Delta\varphi$) at outputs of C4 and C5 couplers (see Fig. 1 b) right). Therefore the phase of the combiner output signal (which is the TapPoint output) related to the input signal source does not change. It is still the same as before the phase drift appeared. That is the basic mechanism of canceling out the main coaxial cable phase drifts at the interferometer link output.

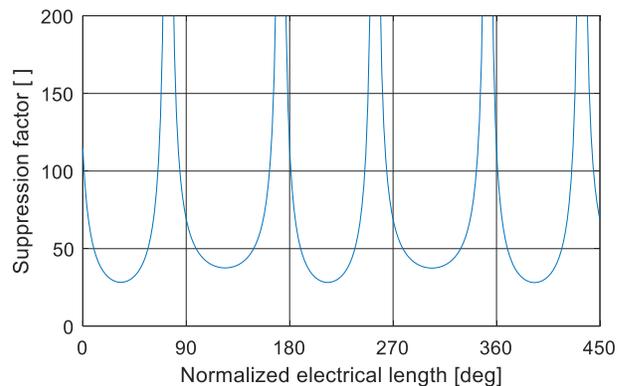

Fig. 2. Basic simulation results of suppression factor

A simulation was performed to plot the SF against the electrical length of the cable between the InCon and TapPoint. Derived plot is shown in Fig. 2. It happens in two ranges per each 180°, but not exactly every each 90°, that the SF value is very large. Unfortunately, the SF peaks are quite narrow. The position and shape of SF peaks and also the bottom level of the SF plot depend on values of the error signals in both C4 and C5 directional couplers. Placing TapPoint operating point within one of those SF peaks is performed by fulfilling the 3rd basic condition, which is in fact a way to minimize the impact of error signals.

## III. Main Error Sources

As it was mentioned above the main issue of the interferometer concept are the error signals that appears in the combined forward and reflected signals at the coupled outputs of C4 and C5. They appear due to a finite isolation of couplers C4 and C5 and RF mismatches at ports of all the signal chain components. The forward signal error in the coupler C4 and the reflected signal error in the coupler C5 are shown in Fig. 3 (green and red arrows). The phase of each error signal is changing in opposite direction than the phase of the coupled signals (Fig. 4). This causes that the resultant phase drifts of the forward and reflected signals that are combined together are not precisely equal ($\Delta\varphi_{FOR} \neq \Delta\varphi_{REF}$) and therefore a phase change of the combined signal may appear. It significantly lowers the drift Suppression Factor value.

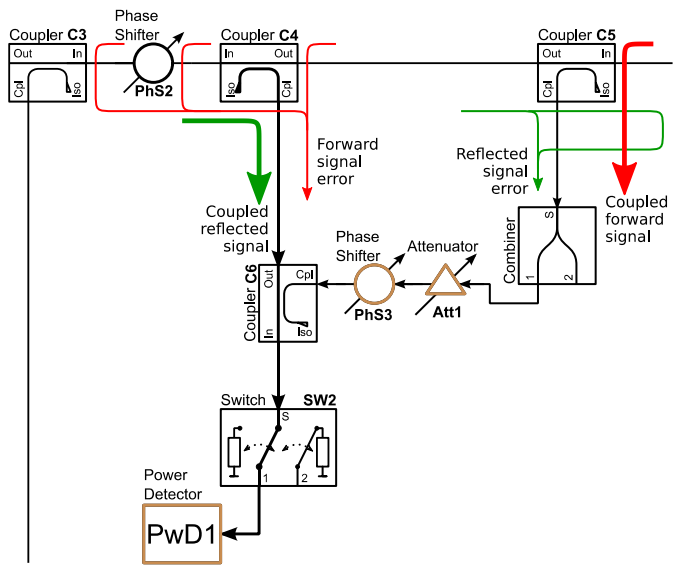

Fig. 3. Errors issue and isolation improvement circuit.

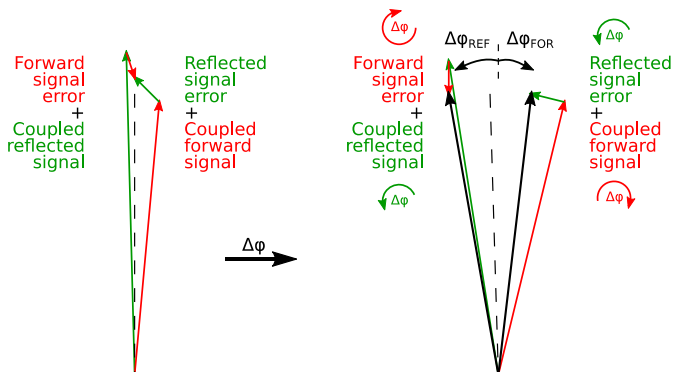

Fig. 4. Example of error and coupled signals behavior on phase drift.

It must be noticed, that the level of the forward signal error (Fig. 3) can be significantly high when compared to the usable reflected signal at the C4 coupler output. This is because the reflected signal must go forth and back through the Cable 2 and all the remaining components including a lossy GND short. Therefore the C4 coupler must be selected with as high isolation as possible. Typical coupler isolation values range between 20 and 30 dB, which was found to be insufficient for this link to achiveve required performance.

## IV. Isolation Improvement

Relatively high level of the forward error signal causes significant errors in the TapPoint adjustment and operation. For this reason the isolation improvement circuit is added (Fig. 3) made of a variable attenuator Att1, a phase shifter PhS3 and a directional coupler C6 used here as a power combiner. The isolation improvement requires turning off the reflected signal at TapPoint input. It is done by the switch SW1 in InCon (Fig. 1) which can be used to terminate the end of main line link with a 50 Ω load. In this state of SW1, the reflected signal disappears and only the forward and forward signal error are present in the TapPoint. In the coupler isolation improvement procedure, the switch SW2 in TapPoint is set to deliver C6 output signal to the power detector PwD1, which is used to measure the resulting error level. The phase shifter PhS3 and the attenuator Att1 are tuned to deliver an additional signal to C6, which is in counterphase to the forward signal error and can reduce the resultant error at C6 output. Both Att1 and PhS3 are adjusted until the level of the error signal at C6 output reaches level of down to 50 dBc below the reflected signal.

## V. Link adjustments

The basic conditions requires measurements of signal parameters (e.g. amplitude and phase) of the forward, reflected and output signals. In the phase averaging reference line concept [4-7] the measurements of signal's phase relationships are done manually at each TapPoint before link operation. The link was disconnected at the right side input of C5 coupler and a 4-port VNA device was used to measure and adjust phase and amplitudes at coupled outputs of the C4 and C5. It is a very laborious procedure and leads to errors from disconnecting and connecting cables at C5 coupler.

We have found out, that it is possible to adjust coupled forward and reflected signal parameters by introducing a C3 directional coupler, which provides a fraction of a forward signal to the TapPoint. Assuming that the adjustment procedure is short enough to neglect the temperature induced phase drifts in the main cable and that the InCon phase lock works properly, the C3 output signal can be treated as phase stable against the signal source (and the GND short) and can be used by the TapPoint as a temporary phase reference for measurements. It is sufficient to measure phase changes of the forward and reflected signals against this temporary reference signal to properly adjust the TapPoint.

## VI. Basic Control Algorithm

Basic controls of the interferometer calibration was mainly implemented in a PC server as a Python script. Few procedures that requires many read and write operations with the RF hardware were implemented in the FPGA I/O board to minimize the time between the signal parameter measurements and to speed the procedures up.

A basic control algorithm is shown in Fig. 5. Briefly it is scanning of the electrical length between the short and the TapPoint C4 and C5 couplers with a calculations of the phase drift suppression factors to find the best operation point (finding one of peaks shown in Fig. 2). The maximum step of the sweep depends on the phase range where the suppression factors are greater than desirable value. Otherwise the algorithm may not find the proper operating point.

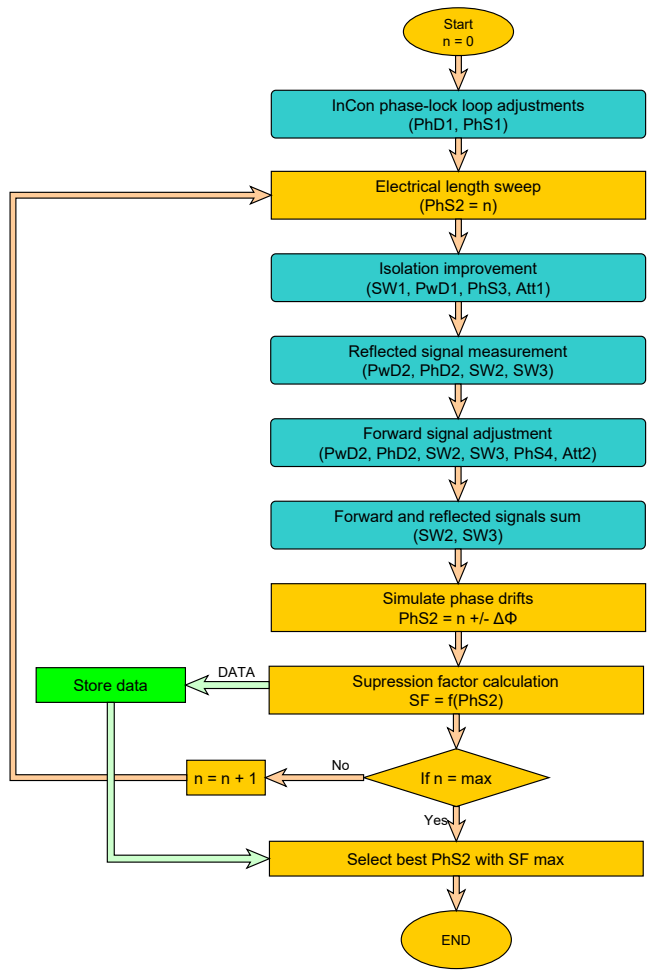

Fig. 5. Basic control algorithm of automatic adjustment

At the beginning the InCon phase-lock loop is run with proper settings of the phase detector PhD1 and the phase shifter PhS1. It is important because the phase-lock loop needs to have enough phase shift margin to compensate full scanning of the automatic adjustment and also to compensate the phase drifts from the Cable 1.

After that the electrical length scanning is started using the main line phase shifter PhS2 in the TapPoint. The consecutive position of the phase shifter is set and the isolation improvement procedure is done to minimize the error on the reflected signal. The procedure is done at each electrical length because the PhS2 regulation changes own matching and transforms the C3 mismatch in different way, so the forward signal error changes too.

Then the reflected signal amplitude and phase are measured with proper configuration of switches SW2 and SW3 that pass only the reflected signal and close the forward signal. Next step change the switches configuration to measure amplitude and phase just of the forward signal. The phase shifter PhS4 and the attenuator Att2 are regulated to get the forward signal equal to the reflected signal measured before. The equalization in the interferometer concept is more accurate because the phase and amplitude are measured directly at the output, so the phase and amplitude unbalances of the combiner are taken into account.

In reality the isolation improvement and the forward signal adjustment to the reflected signal procedures are more complex than in Fig. 5. It is not done in the straight 3 steps. The PhS4 and Att2 regulation causes matching variations at the switch SW3 input, so it has impact on the additional signal generated in the isolation improvement procedure. In fact these steps are done alternately that at the end the isolation improvement is done at final position of the forward signal adjustment.

Later the switches configuration are set that TapPoint combine the forward and reflected signals to get the output signal. The next algorithm step simulate the phase drift in the main line by using the phase shifter PhS2 which is the second feature of the phase shifter. The best would be to simulate the phase drift by a significant value like the real phase drift in the accelerator [8] to have direct answer of the link performance based on the real-like disturbances. The bigger phase drift simulation would also be better to have possibility to measure the suppressed phase drifts at the output significantly above the noise. Unfortunately the phase drift simulation has to be small because the PhS2 regulation changes the forward signal error which has influence on the interferometer calibration. The phase drift at the TapPoint output is then measured and the algorithm calculate the suppression factor for the actual electrical length and store this result. Then the consecutive electrical length point is set and previous steps are done again to the last position of the main line phase shifter PhS2.

At the end of the calibration the stored date are analyzed and the PhS2 position is selected with the best suppression factor.

VII. DESIGNED HARDWARE

The interferometer concept was developed for European XFEL synchronization system. The link components are designed to operate with 1.3 GHz signal which is a main reference signal frequency in the facility. Components used in the hardware design are commonly

The adjustments accuracy in the interferometer link is very important, so stability of environment conditions is very critical. To minimize system own phase and amplitude drifts caused by the humidity variations the RF sub-modules are closed in sealed aluminum enclosures with O-rings. TapPoint module shown in Fig. 6 is an example of that solution.

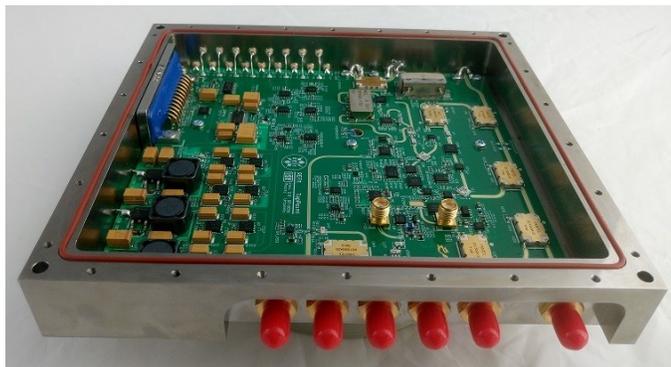

Fig. 6. TapPoint module

To reduce own drifts caused by the temperature changes the temperature of all RF sub-modules is stabilized.

To demonstrate the system concept and make the performance tests a 19" module was built which can be easily installed in the accelerator tunnel. The 19" module is divided into 2 layers to isolate digital electronics from the RF parts. Bottom one contain the FPGA I/O board and other digital boards. The top layer which is shown in Fig. 7 contain InCon and TapPoint modules on the left and right side respectively. The entire top layer is temperature stabilized.

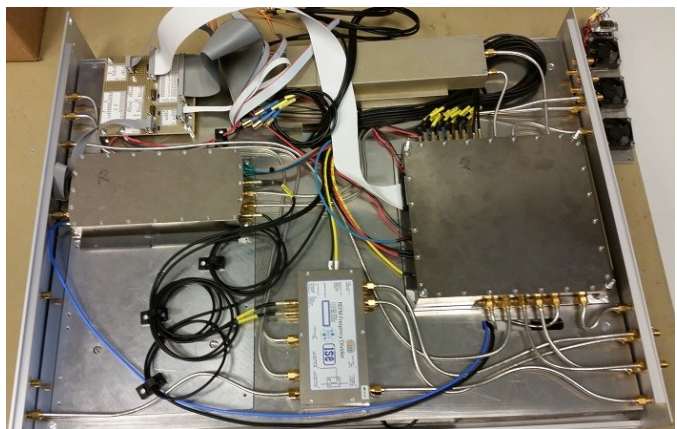

Fig. 7. 19" module with interferometer link

### VIII. INTERFEROMETER TEST RESULTS

The RF interferometer link prototype was tested in the laboratory, what is shown in Fig. 8. The 19" module with the RF sub-modules are placed close to the climate chamber. The long main coaxial cable was located in the climate chamber to get the real cable phase drifts forced by the temperature variations for the performance verification. At the beginning of the test the interferometer calibration was done to set the best operating point. The entire test took more than 80 hours when the temperature was changing in many cycles to obtain similar phase drifts in the cable like in the E-XFEL tunnel.

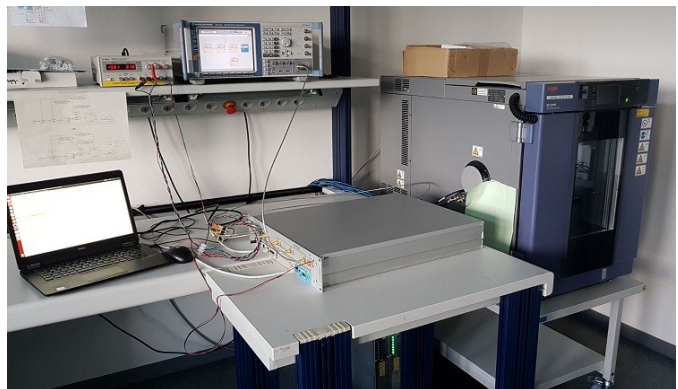

Fig. 8. Interferometer link prototype test stand

Additionally to the interferometer link schematic a directional coupler and 3-way power divider and phase detectors were added for the performance verification of the real suppression factor (Fig. 9). The phase detectors were used to measure the main coaxial cable phase drift and the interferometer link phase drift. The 3-way power divider was used to distribute the reference signal to the RF link prototype and to both phase detectors. The directional coupler was placed in the main line between the coaxial cable and TapPoint for the cable phase drift measurement. All these components were loosely placed at top layer of the 19" module where the temperature was stabilized. This action stabilized the out of loop detectors below 300 fs of phase drift accuracy.

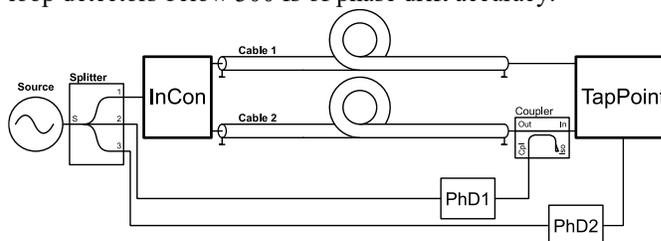

Fig. 9. Interferometer prototype tests schematic.

For the accelerator purpose where the extreme stable time synchronization is required and to become independent of signal frequencies the phase drifts results are represented by time delay drifts. The maximum phase drifts measured in the tunnel were roughly 12.3 ps during the accelerator running up from 2 days downtime [8]. The performance test results are shown in Fig. 10. During the tests the main coaxial cable phase drifts were roughly 10 ps peak-peak, +4.6 ps and -5.7 ps from the operating point. The phase drifts at the interferometer link output are significantly suppressed. They are less than 50 fs peak-peak what is correspond to the suppression factor of 200.

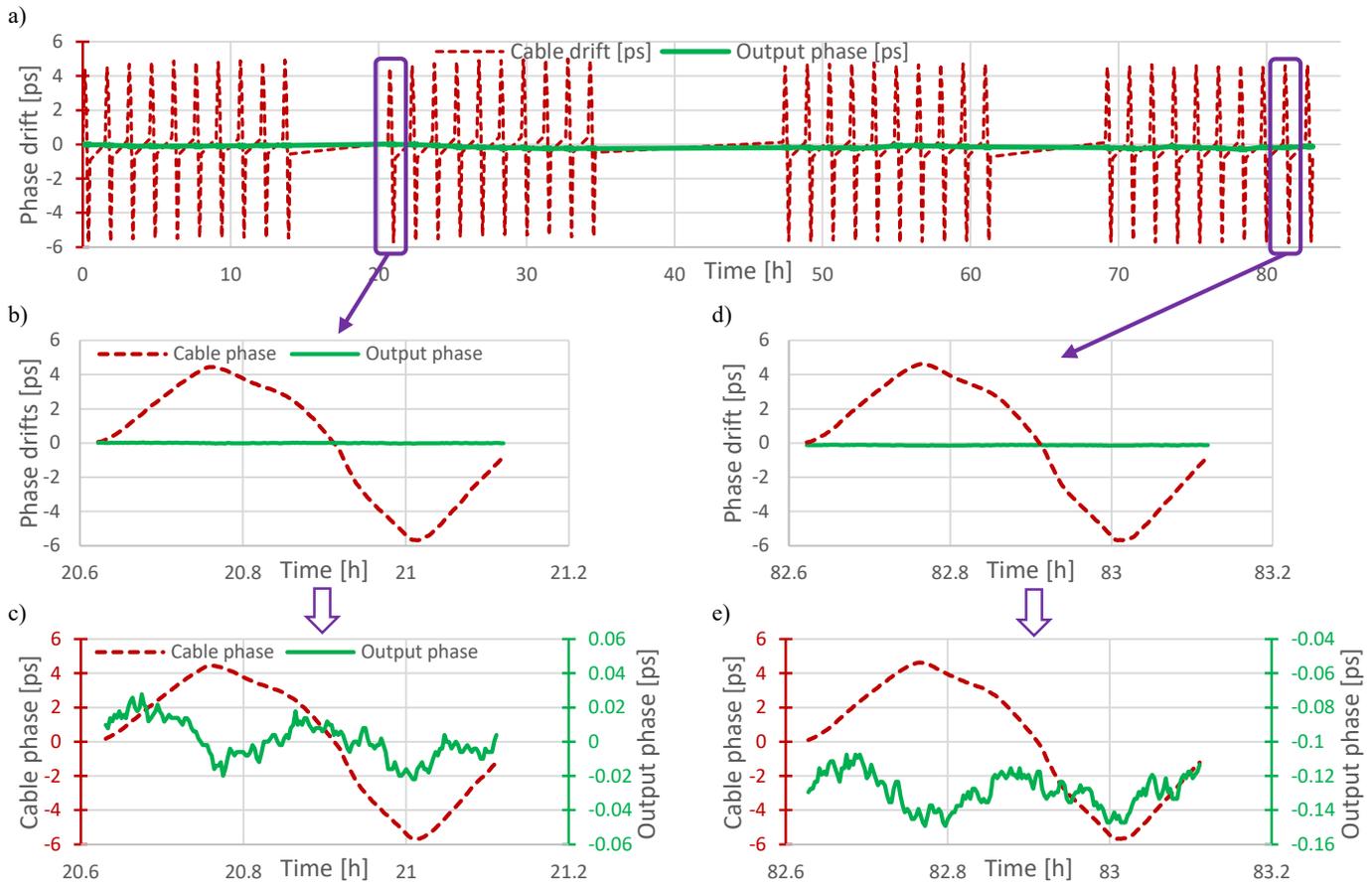

Fig. 10. Interferometer test results: main coaxial cable phase drifts and interferometer output phase drifts vs. time. a) Results over entire 85 h test, b) Results from phase change cycle at 21 h, c) Results from phase change cycle at 21 h with zoomed output phase drifts in, d) Results from the cycle at 83 h, c) Results from the cycle at 83 h with zoomed output phase drifts in.

## IX. Conclusion

The interferometer link prototype shows that the new distribution link concept can be automatically calibrated. Therefore it is more advanced solution compared to the previous concept. It also shows that the level of the E-XFEL phase drifts can be suppressed to the sufficient values.

## References

[1] European XFEL, https://www.xfel.eu/
[2] K. Czuba et al., "Overview of the RF synchronization system for the European XFEL," in *Proceedings of IPAC2013*, Shanghai, China, May 2013, paper WEPME035, pp. 3001–3003. Available: http://accelconf.web.cern.ch/AccelConf/IPAC2013/papers/wepme035.pdf
[3] K.Czuba, D. Sikora, "Temperature stability of coaxial cables," ACTA PHYSICA POLONICA A, Vol. 119, Number 4, 2011, p. 553-557.
[4] E. Cullerton, "1.3 GHz Phase Averaging Reference Line for Fermilab's NML," presented at the 5th Low-Level Radio Frequency Workshop LLRF 2011, Hamburg, Germany, Oct. 17-20, 2011. Available: https://indico.desy.de/indico/event/3391/session/6/contribution/27/material/slides/0.pptx
[5] E. Cullerton, B. Chase, "The Design and Performance of the Fermilab ASTA Phase Averaging 1300 MHz Phase Reference Line," presented at the 6th Low-Level Radio Frequency Workshop LLRF 2013, California, USA, Oct. 1-4, 2013. Available: https://conferences.lbl.gov/event/27/session/23/contribution/79/material/poster/0.pdf
[6] B. Chase and E. Cullerton, "1.3 GHz Phase Averaging Reference Line for Fermilab's NML ASTA Program", Beams-doc-3806-v2. Available: http://beamdocs.fnal.gov/AD/DocDB/0038/003806/002/Reference%20Line_v3.docx
[7] L. Doolittle et al., "The LCLS-II LLRF System," in *Proceedings of IPAC2015*, Richmond, VA, USA, 2015, paper MOPWI021, pp. 1195-1197. Available: http://accelconf.web.cern.ch/AccelConf/IPAC2015/papers/mopwi021.pdf
[8] T. Lamb et al., "Laser-to-RF Synchronization with Femtosecond Precision," in *Proceedings of FEL2017*, Santa Fe, NM, USA, 2017, paper WEB04, pp. 399-402. Available: http://fel2017.jacow.de/papers/web04.pdf